\documentclass[prl, aps, twocolumn, groupedaddress, superscriptaddress]{revtex4}
\usepackage{epsfig}
\usepackage{slashed}
\usepackage{graphicx}
\usepackage[usenames, dvipsnames]{color} 
\usepackage{psfrag}
\usepackage{graphics}
\usepackage{hyperref}

\begin{document}

\title{Two-stream instability in quasi one-dimensional Bose-Einstein condensates}

\author{H. Ter\c{c}as}
\email{htercas@cfif.ist.utl.pt}
\affiliation{CFIF, Instituto Superior T\'{e}cnico, Av. Rovisco Pais 1, 1049-001 Lisboa, Portugal}
\author{J. T. Mendon\c{c}a}
\email{titomend@ist.utl.pt}
\affiliation{CFIF, Instituto Superior T\'{e}cnico, Av. Rovisco Pais 1, 1049-001 Lisboa, Portugal}
\affiliation{IPFN, Instituto Superior T\'{e}cnico, Av. Rovisco Pais 1, 1049-001 Lisboa, Portugal}
\author{G. R. M. Robb}
\affiliation{SUPA, Department of Physics, University of Strathclyde, Glasgow G4 0NG, United Kingdom}

\begin{abstract}
We apply a kinetic model to the predict the existence of a new instability mechanism in elongated Bose-Einstein condensates. Our kinetic description, based on the Wigner formalism, is employed to highlight the existence of unstable Bogoliubov waves that may be excited in the counter-propagation configuration. We identify a dimensionless parameter, the Mach number at $T=0$, that tunes different regimes of stability. We also estimate the magnitude of the main parameters at which two-stream instability is expected to be observed under typical experimental conditions.  
\end{abstract}

\maketitle

Several aspects of BEC  physics have been explored in recent years, with particular relevance to collective processes, such as collective atomic recoil lasing (CARL) \cite{piovella, fallani}, the excitation of Bogoliubov waves in elongated condensates \cite{stamper, vogels, ozeri}, Bloch oscillations in spatially periodic potential induced by a far-off-resonance laser field \cite{berg} and the observation of quantum fluctuations and entanglement \cite{piovella2, dunningham, sorensen, hines}. Attention to the physics of BECs has also been paid in astrophysics, since some recent theories suggest BECs as good candidates for dark matter \cite{bohmer}. Recent wave kinetic models have been proposed, in both classical and quantum regimes, to properly describe many cooperative phenomena in BECs, like Bogoliubov waves, wakefields and instabilities \cite{mendonca1, mendonca2, bonifacio, konotop}. Exact kinetic equations also seem to constitute an essential tool in the description of thermodynamics of condensation. For example, Gardiner et al have applied kinetic methods based on the Wigner function to describe the formation of a BEC in the presence of a thermal bath \cite{gardiner}. 

In this Brief Report, we apply a quantum kinetic model to study the dynamical instability of Bogoliubov waves in two counter propagating BECs. The occurrence of instabilities in BECs has been discussed in several works, ranging from instabilities of the superflow in optical lattices \cite{Wu} and dynamical instabilities of rotating BECs \cite{bijnen} to vortex formation mechanisms \cite{parker}. Recent experimental and theoretical works on collisions of BECs \cite{kozuma, norrie}, reporting on the scattering of atoms and quantum turbulence, pave the stage where such instabilities may take place. Furthermore, in experiments performed on storage ring traps for ultracold atoms \cite{houston}, where BECs can be split into two components which counter-propagate round the ring, the study of the stability criteria manifestly becomes an issue of major relevance.  We anticipate that, in quasi one-dimensional systems, the splitting of the condensate into two parts that are set to move against each other, the Bogoliubov waves can become dynamically unstable. Such an instability should be referred as a ''two-stream'' instability, in complete analogy with the well known phenomenon in plasma physics \cite{nicholson}. This dynamical instability drives an exponential increase of the amplitudes of the fluctuations in the condensate. Such a growth induces the dephasing of the condensate, transferring its translational kinetic energy to collective and single particle excitations (phonons). We observe two different regimes of instability, occurring in both subsonic and supersonic regimes. 


In the spirit of mean-field theory, the collective field operator $\hat \Psi(\mathbf{r},t)=\Phi(\mathbf{r},t)+\hat \psi(\mathbf{r},t)$ of the Bose gas can be separated into a condensate $c$-number wave function $\Phi(\mathbf{r},t)=\langle \Psi(\mathbf{r},t)\rangle$, where $\langle\cdot\rangle$ stands for the expectation value in the ground state, and its quantum fluctuations $\hat \psi(\mathbf{r},t)$. The condensed field verifies the normalization condition $\int \Phi(\mathbf{r},t)^* \Phi(\mathbf{r},t)d\mathbf{r}=N$, where $N$ is the total number of condensed atoms. The dynamics of the BEC wave function is governed by the celebrated GP equation  

\begin{equation}
i\hbar\frac{\partial\Phi}{\partial t}=-\frac{\hbar^2}{2m}\nabla^2\Phi+(V_0+V_{SC})\Phi,
\label{eq:GP}
\end{equation}
where $V_{0}=V_0(\mathbf{r})=m/2(\omega_{\perp}^2r_{\perp}^2+\omega_{z}^2z^2)$ represents the asymmetric confining potential, $m$ is the single atom mass, $r_{\perp}=(x^2+y^2)^{1/2}$ and $z$ represent the transversal and longitudinal directions, respectively, and $V_{SC}=g\vert\Phi(\mathbf{r},t)\vert^2$ is the nonlinear self-consistent potential. Here, $g=4\pi\hbar^2 a/m$ defines the coupling constant or interaction strength and $a$ represents the $s$-wave scattering length. The main ingredient for the derivation of a quantum kinetic equation is the two-point correlation function associated to the condensed wave function 

\begin{equation}
C(\mathbf{s},\tau)=\Phi(\mathbf{r}-\mathbf{s}/2,t-\tau/2)\Phi^*(\mathbf{r}+\mathbf{s}/2,t+\tau/2),
\label{eq:corr}
\end{equation}
where $\mathbf{s}=\mathbf{r}_1-\mathbf{r}_2$, $\mathbf{r}=(\mathbf{r}_1+\mathbf{r}_2)/2$, $t=(t_1+t_2)/2$ and $\tau=t_1-t_2$. By performing a double Fourier transform of the relative position ($\mathbf{s}$) and time ($\tau$) variables, we define the Wigner function

\begin{equation}
W(\mathbf{r},t;\omega,\mathbf{k})=\int d\mathbf{s}\int d\tau C(\mathbf{s},\tau) \exp(i\omega\tau -i\mathbf{k}\cdot\mathbf{s}),
\label{eq:wigner1}
\end{equation}      
where $\omega$ and $\mathbf{k}$ represent the frequency and the momentum of the BEC excitation, respectively. Combining the GP equation (\ref{eq:GP}) with Eq. (\ref{eq:wigner1}), it is possible to derive the Wigner-Moyal equation \cite{mendonca3}
\begin{equation}
\begin{array}{cc}
\displaystyle{i\hbar\left(\frac{\partial}{\partial t}+\frac{\hbar}{m}\mathbf{k}\cdot\mathbf{\nabla}\right)W
=\int \frac{d\mathbf{q}}{(2\pi)^3} \int \frac{d\Omega}{2\pi}V(\mathbf{q},\Omega)}\\\\
\displaystyle{\times\left[ W_--W_+\right] \exp(-i\Omega t +i\mathbf{q}\cdot\mathbf{r}),}
\end{array}
\label{eq:wigner2a}
\end{equation}
where $W_{\pm}=W(\omega\pm\Omega/2,\mathbf{k}\pm\mathbf{q}/2)$ and 

\begin{equation}
V(\mathbf{q},\Omega)=\int d\mathbf{r}\int dt V(\mathbf{r},t)\exp(-i\mathbf{q}\cdot\mathbf{r}+i\Omega t)
\label{eq:wigner2b}
\end{equation}
stands for the double Fourier transform of the total potential $V=V_0+V_{SC}$. We linearize the system by introducing a perturbation around its equilibrium configurations, such that $W_\pm=W_{0\pm}+\tilde W_{\pm}, \quad V=V_0+\tilde V$, where $\tilde W_\pm=\tilde W_\pm(\mathbf{q},\Omega) \exp(i\mathbf{q}\cdot\mathbf{r}-i\Omega t)$ and $\tilde V=\tilde V(\mathbf{q},\Omega)\exp(i\mathbf{q}\cdot\mathbf{r}-i\Omega t)$ represents the single Fourier components of the wave. We restrict our discussion to single Fourier components, since we are looking for the stability of a single mode. The same formalism can be easily extended to a broadband fluctuation spectrum, which is not the aim of the present work. Assuming that the confining potential is static, $V_0=V_0(\mathbf{r})$, the fluctuations will be only rooted in the nonlinear mean-field potential. This statement restricts our discussion to the case of dynamical instabilities, which has no dependence on the shape of the external potential, in contrast to the case of energetic instabilities \cite{menotti}. The Fourier component of the perturbed potential is therefore given by the convolution $\tilde V(\mathbf{q},\Omega)=g(\Phi^**\Phi)(\mathbf{q},\Omega)$. We can easily obtain the linearized version of Eq. (\ref{eq:wigner2a})

\begin{equation}
\begin{array}{lr}
\displaystyle{\left(\frac{\partial}{\partial t}+\mathbf{v}\cdot\mathbf{\nabla}\right)\tilde W }\\\\

\displaystyle{=\frac{g}{i\hbar}\int\frac{d\mathbf{q}}{(2\pi)^3}\int\frac{d\Omega}{2\pi}\tilde V(\mathbf{q},\Omega)\left[W_{0+}-W_{0-}\right]\exp{i\theta}},
\end{array}
\label{eq:wigner3}
\end{equation}
where $\mathbf{v}=\hbar \mathbf{k}/m$ represents the velocity field of the BEC and $\theta=\mathbf{q}\cdot\mathbf{r}-\Omega t$ is the wave phase. \par
In order to obtain very elongated BECs, one requires an asymmetric trapping potential $V_{0}(\mathbf{r})=V_{\perp}(\mathbf{\rho})+V_{z}(z)$, such that the condition $\omega_{\perp}=(\omega_{x}^2+\omega_{y}^2)^{1/2}\gg\omega_{z}$ is satisfied. A typical experiment with $^{87}$Rb atoms contains $N\sim 10^5$ atoms confined in an asymmetric harmonic trap with radial and axial trapping frequencies of $\omega_{\perp}/2\pi\sim 200$ Hz and $\omega_{z}/2\pi\sim 20$ Hz, respectively \cite{steinhauer}. The radial and axial Thomas-Fermi lengths of the BEC are $R\approx 3.1$ $\mu$m and $Z\approx 27.1$ $\mu$m. Under these conditions, we can assume that the BEC is quasi-1D. Assuming a static density profile along the transversal direction, the perturbations of the self-consistent field along the $\hat\mathbf{z}$-axis are given by $\tilde V(z,t)=\tilde V(q_{z},\Omega)\exp(iq_zz-i\Omega t)$, which together with Eq. (\ref{eq:wigner3}) yields

\begin{equation}
\tilde W(z,t;\omega,\mathbf{k})=\frac{g}{\hbar}\tilde V(z,t)\frac{W_{0-}-W_{0+}}{\Omega-q_{z}v_z}.
\label{eq:wigner4} 
\end{equation}
Integration of Eq. (\ref{eq:wigner4}) over the BEC spectrum $(\mathbf{k},\omega)$, together with the identity
\begin{equation}
\tilde V(z,t) = \int\int \frac{d\mathbf{k}}{(2\pi)^3}\frac{d\omega}{2\pi} \tilde W(z,t;\mathbf{k},\omega)
\label{eq:poisson}
\end{equation}
and considering that the condensate spectrum can be written in the form $\omega=\omega(\mathbf{k})$, which allows the factorization $\tilde W(z,t;\omega,\mathbf{k})=2\pi \tilde W(z,t;\mathbf{k})\delta(\omega-\omega(\mathbf{k}))$, yields the following kinetic dispersion relation

\begin{equation}
1-\frac{g}{\hbar}\int \frac{d\mathbf{k}}{(2\pi)^3}\frac{W_0(z;\mathbf{k}-\mathbf{q}/2)-W_0(z;\mathbf{k}+\mathbf{q}/2)}{\Omega-\mathbf{q}\cdot \mathbf{v}}=0.
\label{eq:wigner5}
\end{equation}
In the one-dimensional approximation, the waves can not propagate in the radial direction, and we can therefore factorize the equilibrium Wigner function into its longitudinal and perpendicular components, such that $W_0(z;\mathbf{k})=(2\pi)^2W_0(z;k_z)\delta (\mathbf{k}_\perp)$. Dropping the subscript $z$, for the sake of simplicity, we can finally write down the dispersion relation for the longitudinal oscillations

\begin{equation}
1-\frac{g}{\hbar}\int \frac{d k}{2\pi}W_0(z;k)\left[\frac{1}{\Omega_+-qv}-\frac{1}{\Omega_--qv}\right]=0,
\label{eq:wigner6}
\end{equation}
where $\Omega_\pm=\Omega\pm\hbar q^{2}/2m$. One can easily obtain the usual excitation spectrum at $T=0$, by simply setting the corresponding equilibrium profile $W_0(z;k_z)=2\pi  n_0  \delta(k_z-k_0)$, which yields $(\Omega-qv_0)^2=u_B^2q^2+\hbar^2q^4/4m^2$, where $v_{0}=\hbar k_{0}/m$ is the stream speed, $u_B=\sqrt{g n_0 /m}$ represents the Bogoliubov sound speed and $n_{0}=\langle  n_0(\mathbf{r})\rangle $ is the average background density. We immediately conclude that our kinetic approach is totally equivalent to the usual Bogoliubov-de Gennes method. Equation \ref{eq:wigner6} shows that these (doppler shifted) second-sound waves are dynamically stable, since $\Omega$ is always real. One should notice that the dynamical instability has no relation with the Landau criterion of superfluidity (energetic instability), which strongly depends on the external potential. Menotti et al have explicitly shown the difference between energetic and dynamical instability in the case of a BEC confined in an optical lattice. \cite{menotti}. \par
\begin{figure}
\includegraphics{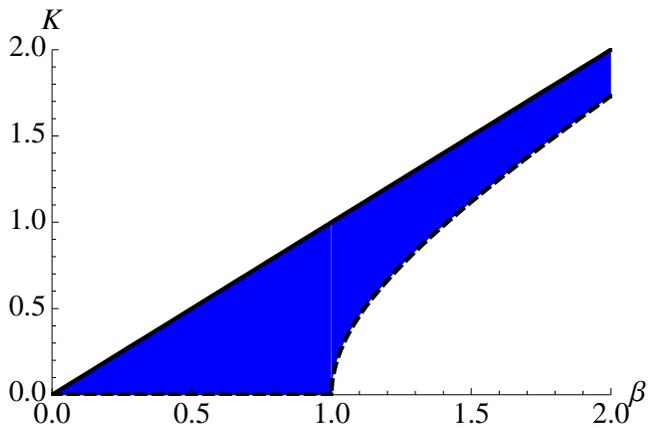}
\caption{(Color Online) Stability diagram for the two-stream BEC. The shadowed area represents the set of parameters for which the Bogoliubov waves are unstable. The upper curve corresponds to $K=\beta$. The dashed lower curve corresponds to the case $K=\sqrt{\beta^2-1}$}
\label{fig1}
\end{figure}
\begin{figure}
\includegraphics{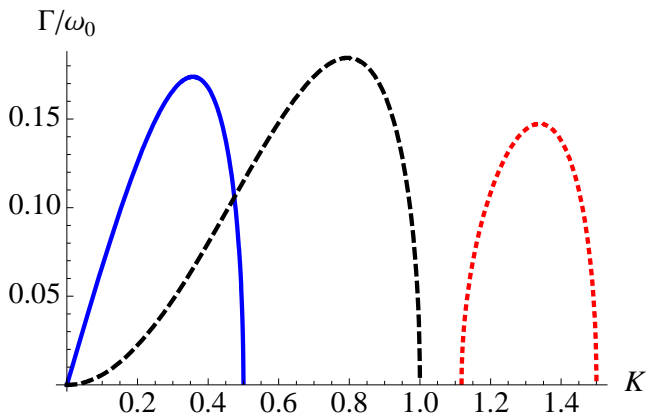}
\caption{(Color Online) Normalized wave growth rate $\tilde \Gamma=\mbox{Im}(\tilde \Omega)/\omega_{0}$ for different values of $\beta$. Blue full line, $\beta=0.5$, black dashed line, $\beta=1.0$, and red dotted line, $\beta=1.5$. The maximum growth rate, corresponding to the most unstable mode, occurs for $\beta=1$ and $K\approx 0.8$.}
\label{fig2}
\end{figure}

In the counter-propagation configuration, a discussion about the stability criteria of the excitations deserve some attention. In order to give a first insight to the problem,  we assume a quasi-equilibrium profile describing two incoherent BEC beams, which is approximately given by $W_0(\mathbf{r};k)=\pi n_0 \left[\delta(k-k_0)+\delta(k+k_0)\right]$, where $ n_{0} $ is the mean density. This assumption is made by noticing that, at $T=0$, the rms velocity $v_{rms}=\hbar\sqrt{\langle k^2\rangle}/m$ of each beam is very small, such that $q v_{rms} \ll \Omega$, which is satisfied for modes such that $1/q$ is much less than the Thomas-Fermi length $Z= \sqrt{m\omega_{z}/\hbar}$. We test the consistence of this approximation at the end of this work. Under these conditions, the dispersion relation for the sound waves in the two-stream configuration is finally given by

\begin{equation}
1-\frac{K^2}{2}\left\{\frac{1}{\beta^2(\tilde\Omega+K)^2-K^4}+\frac{1}{\beta^2(\tilde \Omega-K)^2-K^4}\right\}=0,
\label{eq:bog2}
\end{equation} 
where we have defined the normalized quantities $\tilde \Omega=\Omega/\omega_{0}$, $K=\hbar q/mu_{B}$, and the dimensionless parameter $\beta=v_{0}/u_{B}$. Here, $\omega_{0}=u_{B}k_{0}$ represents the 'resonance' frequency. The parameter $\beta$ defines a sonic number, the Mach number at $T=0$. It measures if the beam flow is either subsonic ($\beta < 1$) or supersonic ($\beta > 1$). The factorization of Eq.(\ref{eq:bog2}) provides two branches

\begin{equation}
\tilde\Omega^2_{\pm}=\frac{K^2}{2\beta^2}\left[1+ 2\beta^2+2K^2\pm \sqrt{1+8\beta^2 + 16\beta^{2} K^2})\right],
\label{eq:bog3}
\end{equation}
one of which $(\tilde\Omega^2_{+})$ is always positive and describes stable oscillations. However, the solution $\tilde\Omega_{-}^2$ is not positive definite and negative solutions are found for the modes $K$ that verify the condition
$1 + 2K^2 +2\beta^2\leq (1 +8 \beta^2 + 16\beta^2 K^2)^{1/2}$. This condition defines two dynamically unstable regions in the $(K, \beta)$ plan, as illustrated in Fig. (\ref{fig1}). In the subsonic regime, $\beta < 1$, all the modes satisfying $K < \beta$ are unstable; in the supersonic regime, $\beta>1$, unstable modes are obtained for $\sqrt{\beta^2-1} < K < \beta$. The representation of the dimensionless growth rate $\tilde \Gamma(K, \beta)=\Gamma/\omega_{0}$, where $\Gamma=\Im(\Omega)$, shows that such unstable oscillations occur in both supersonic and subsonic regimes (see Fig.(\ref{fig2})). In the subsonic regime, all wave modes $K$ are unstable, until the cut-off given by the condition $K=\beta$. This means that all $q$-modes are unstable up to the cut-off wave vector $q_c=k_{0}$, when this mode has the same energy of the beam. In that case, all the wave energy is transferred to the condensate, and the instability vanishes. In the supersonic regime, $\beta>1$, the picture changes. First, to become unstable, the phonon must have a momentum higher than the resonant value $q_r=\gamma mu_{B}/\hbar$, where $\gamma=[(v_{0}/u_{B})^{2}-1]^{1/2}$. It means that the phonon must be resonant with that of the supersonic beam, so energy can be transferred from the condensate to the wave. Our calculations also indicate that the increasing of $\beta$ increases both the value of the cut-off mode $q_c$ and the value of the resonant mode $q_{r}$. In the subsonic (supersonic) regime, the value of the maximum growth rate increases (decreases) as $\beta$ increases. Therefore, the most unstable mode is obtained for $\beta=1$ and corresponds to a maximum growth rate of $\Gamma_{max}\approx 0.185 \omega_{0}$, relative to the mode $q_{max}\approx 0.795 q_{B}$, where $q_{B}=mu_{B}/\hbar$. For the typical experimental conditions performed with $^{87}$Rb \cite{steinhauer}, the average density is $n_{0}\sim 10^{11}$cm$^{-3}$ and the $s$-wave scattering length is $a\sim 1$ nm, as it is known for alkali metals \cite{bloch}, which provides a Bogoliubov speed of the range of $u_{B}\sim 10$ cm/s. Preparing the condensate beams at $\beta=1$, we expect to observe a maximum growth period of $\tau=2\pi/\Gamma_{max}\sim 3$ ms. The typical wave length for which two-stream instability is expected to occur is at the order of $\lambda=2\pi/q_{B}\sim 1~\mu$m, which is consistent with the quasi-one dimensional approximation we have performed. One should also notice that the validity criterion for the Bogoliubov theory $\sqrt{n_{0}a^3}\sim 10^{-3}\ll1$ is also satisfied for the present experimental conditions. A final remark about the validity of the present results is in order. First, it is well known that under conditions of reduced effective dimensionality, phase fluctuations play an important role (our present discussion is limited to weakly interacting systems whose kinematics is low dimensional, but the scattering can still be seen as a three dimensional process, which is far away from the Tonks-Gireardeau regime). For such a system, below the degeneracy temperature $T_{d}=N\hbar\omega_{t}/k_{B}$, where $\omega_{t}=\sqrt{\omega_{\perp}^2+\omega_{z}^2}$, (the low-dimensional analogue of the critical temperature), one can define a second characteristic temperature, the 'phase fluctuation' temperature $T_{ph}=T_{d}\hbar\omega_{t}/\mu$, where $\mu$ is the Thomas-Fermi chemical potential \cite{petrov}. At $T\ll T_{ph}$, the condensate phase fluctuates at scales much smaller than the Thomas-Fermi length, which correspond to the range of temperatures where the two-stream instability mechanism presented here may be observed.  Second, in the presence of a thermal gas, a more detailed calculation should include the dynamics of the decoherence and the respective equilibrium distribution function would correspond to that of a low-dimensional matter wave interferometer \cite{burkov}. A simple model for the condensate at finite temperature introduces a broadening of the equilibirum profile \cite{shukla}, and therefore some wave modes are Landau damped, which suggests that our results, based on the two-independent beams configuration, correspond to the ''best case scenarium'' for this dynamical instability. \par

In conclusion, we have theoretically established the criteria for dynamical instability of Bogoliubov waves in BECs prepared in the two-stream configuration. Threshold conditions for the occurrence of unstable regimes, and the corresponding growth rates were derived. The conditions for the two-stream instability change in the subsonic and supersonic limits, where in the later the instability can not be excited by an arbitrarily small wave vector. The maximum growth rate of the dynamical instability is observed when the nominal velocity of the wave packets equals the sound speed in the condensate. The authors consider that the features of the instability mechanism presented here can eventually provide the basis for future experiments. \par
This work was partially supported by Funda\c c\~ao para a Ci\^encia e Tecnologia (FCT-Portugal), through the grant number SFRH/BD/37452/2007. G.R. would like to acknowledge support from the Leverhulme Trust via research grant F/00273/I.

\end{document}